\documentclass[aps]{revtex4}
\usepackage{graphicx}

\begin{document}
\title{Quantum pure-state identification}
\author{A. Hayashi, M. Horibe, and T. Hashimoto}
\address{Department of Applied Physics\\
           University of Fukui, Fukui 910-8507, Japan}
%\draft

\begin{abstract}
We address a problem of identifying a given pure state with one of 
two reference pure states, when no classical knowledge on the reference 
states is given, but a certain number of copies of them are available.
We assume the input state is guaranteed to be either one of the two reference 
states. This problem, which we call quantum pure state identification, is a 
natural generalization of the standard state discrimination problem.
The two reference states are assumed to be independently distributed in a 
unitary invariant way in the whole state space. 
We give a complete solution for the averaged maximal success probability of 
this problem for an arbitrary number of copies of the reference states 
in general dimension. 
It is explicitly shown that the obtained mean identification probability 
approaches the mean discrimination probability as the number of the 
reference copies goes to infinity. 
\end{abstract}

\maketitle

\newcommand{\ket}[1]{|\,#1\,\rangle}
\newcommand{\bra}[1]{\langle\,#1\,|}
\newcommand{\braket}[2]{\langle\,#1\,|\,#2\,\rangle}
\newcommand{\bold}[1]{\mbox{\boldmath $#1$}}
\newcommand{\sbold}[1]{\mbox{\boldmath ${\scriptstyle #1}$}}
\newcommand{\tr}[1]{{\rm tr}\left[#1\right]}
\newcommand{\BC}{{\bold{C}}}
\newcommand{\CS}{{\cal S}}
\newcommand{\CM}{{\cal M}}
\newcommand{\CA}{{\cal A}}
\newcommand{\CP}{{\cal P}}

%%%%%%%%%%%%%%%%%%%%%%%%%%%%%%%%%%%%%%%%%%%%%%%%%%%%%%%%%%

\section{Introduction}
Suppose we are given an unknown quantum pure state $\rho$ on a $d$ dimensional 
vector space $\BC^d$. 
We know that the input state $\rho$ is either one of two reference states 
$\rho_1$ and $\rho_2$, each also being a pure state on $\BC^d$. 
What is the best strategy to identify the input state with one of the two 
reference states?

We can consider two cases depending on what kind of information on 
the reference states is available. 
In the first case, we are given complete classical knowledge on the 
reference states $\rho_1$ and $\rho_2$. 
This is the problem of quantum state discrimination, which was solved by 
Helstrom \cite{Helstrom76}.

On the other hand, we can also consider the case where only a certain number 
$(N)$ of copies of $\rho_1$ and $\rho_2$ are given, with no classical 
knowledge on them available. 
In this case, we could obtain only limited classical information on the 
reference states, since the no-cloning theorem \cite{Wootters82} does not 
allow us to increase the number of copies of the reference states.
The best we can do is that we perform a POVM measurement on the total state  
$\rho \otimes \rho_1^{\otimes N} \otimes \rho_2^{\otimes N}$ and try to 
identify the input state $\rho$ with one of the reference states $\rho_1$ 
and $\rho_2$.
In this paper, this problem is called "quantum state identification".
If the number of copies $N$ is infinite, the problem is reduced to quantum 
state discrimination. This is because we can always obtain complete classical 
knowledge of a quantum state if we have infinitely many copies of 
the state.
Thus quantum state identification is a natural generalization 
of quantum state discrimination, which is one of the most fundamental problems 
in quantum information theory.

In the case of qubit ($d=2$), similar problems but in different setups have 
already been studied \cite{Sasaki01,Sasaki02,Bergou05}. 
Sasaki {\it et al}. studied quantum matching problem \cite{Sasaki02}, where a 
certain number of copies of input and reference states are given according to 
a known {\it a priori} independent probability.  
Using the fidelity as a figure of 
merit, they determined the optimal fidelity in the case where the input and 
reference qubits are independently distributed on a great circle of the Bloch 
sphere (see also Ref.~\cite{Sasaki01}).
Bergou {\it et al}. \cite{Bergou05} recently discussed a generalization of 
unambiguous discrimination problem \cite{Ivanovic87,Dieks88,Peres88}. 
In this generalization we are given three qubit systems, 
where one of the three qubits, input system, is guaranteed to be prepared in 
one of the states of the other two reference qubits with some occurrence 
probability. 
Without any classical knowledge on the reference states, we are to tell 
unambiguously in which of these reference states the input qubit is prepared. 
In Ref.~\cite{Bergou05} the optimal success probability was derived as a 
function of the occurrence probability of the two cases.

In this paper, we assume that the input state $\rho$ is guaranteed to be 
prepared in one of the two reference states $\rho_1$ and $\rho_2$ like in 
the original discrimination problems and the generalization of unambiguous 
discrimination discussed in Ref.~\cite{Bergou05}.
The two reference states are assumed to be independently distributed on 
the whole $d$ dimensional pure state space $\BC^d$ in a 
unitary invariant way, which will be precisely defined in the next section. 
No classical knowledge on the reference states is available, but only a 
certain number (N) of their copies are given. 
Our task is to successfully identify the input state with one of the reference 
states. 

We will give a complete solution of optimal strategies and the 
mean success probability as a function of the number of copies $N$ of 
the reference states and the dimension $d$ of the state space. 
We also study the large $N$ limit of the mean identification probability 
and verify that it approaches the mean discrimination probability as the 
number of copies $N$ goes to infinity.

%%%%%%%%%%%%%%%%%%%%%%%%%%%%%%%%%%%%%%%%%%%%%%%%%%%%%%%%%%%%%%%%%%%%

\section{Averaged maximum probability of pure state discrimination}
First we calculate the averaged maximum discrimination probability in order to 
compare it with the maximum identification probability derived later.
The success probability of discrimination between two known pure states 
$\rho_1$ and $\rho_2$ is given by
\begin{eqnarray}
  p(\rho_1,\rho_2) = \frac{1}{2} \sum_{a=1}^2 \tr{E_a \rho_a},
\end{eqnarray}
where the occurrence probabilities of $\rho_1$ and $\rho_2$ are assumed to be 
equal.  
Here $\{E_1,E_2\}$ is a set of POVM and with outcome labeled by $a(=1,2)$ 
the input state is guessed to be $\rho_a$. The POVM may depend on the 
reference states $\rho_1$ and $\rho_2$, since we have the classical knowledge 
on $\rho_a$ in this discrimination problem.

For a given set of pure states $\rho_1$ and $\rho_2$, the maximum 
success probability is expressed in terms of the trace distance 
between the two states as
\begin{eqnarray}
  p_{\max}(\rho_1,\rho_2) =
           \frac{1}{2} \left(
           1 + D(\rho_1,\rho_2) \right),
\end{eqnarray}
where $D(\rho_1,\rho_2) \equiv \frac{1}{2}{\rm tr}|\rho_1-\rho_2|$ 
\cite{Helstrom76}. 
We note that the trace distance between two pure states is given by 
\begin{eqnarray}
   D(\rho_1,\rho_2) = \sqrt{1-\tr{\rho_1\rho_2}}.
\end{eqnarray}

Now we specify the distribution by which 
the reference states $\rho_1$ and $\rho_2$ are chosen. 
We assume that the two states are chosen independently and uniformly 
from the pure state space on $\BC^d$ in a unitary invariant way.
Denoting the average over this distribution by $<\cdots>$, 
we define the averaged maximal success probability of state discrimination 
to be $p_{\max}(d) = < p_{\max}(\rho_1,\rho_2) >$.

More precisely, the average is defined as follows.
Let us expand a normalized pure state $\ket{\phi} \in \BC^d$ by an 
orthonormal reference 
base $\{\ket{i}\}_{i=1}^d$ as $\ket{\phi} = \sum_i c_i \ket{i}$ with 
$\sum_i |c_i|^2 = 1$.
The distribution of $\ket{\phi}$ is assumed to be given by 
the uniform 
distribution of $\{x_i \equiv \Re c_i, y_i \equiv \Im c_i \}_{i=1}^d$ 
on the $2d-1$ dimensional hyper sphere. 

When averaging the distance $D(\rho_1,\rho_2)$, we can assume the state 
$\rho_1$ is given by $\ket{1}\bra{1}$, where $\ket{1}$ is one of the vectors 
in the reference base,  
which is justified by the unitary invariance of the distance $D$ and the 
distribution of states.
 
Expanding the state $\rho_2$ in the reference base with coefficients $c_i$, 
we can express the averaged distance as
\begin{eqnarray}
  <D(\rho_1,\rho_2)> = \frac{ \int\! dcdc^+ \sqrt{1-|c_1|^2} }
                            { \int\! dcdc^+ 1},
\end{eqnarray}
where the integration measure corresponding to the uniform distribution 
on the hyper sphere is given by
\begin{eqnarray}
   \int\! dcdc^+ \equiv \int_{-\infty}^{+\infty} \prod_{i=1}^d
     (dx_i dy_i) \delta(\sum_i(x_i^2+y_i^2) -1).  \label{measure}
\end{eqnarray}

Performing the above integrations, we find that the average distance takes 
the form 
\begin{eqnarray}
     <D(\rho_1,\rho_2)> = \frac{d-1}{d-\frac{1}{2}},
\end{eqnarray}
which gives the averaged maximum success probability of discrimination to be 
\begin{eqnarray}
   p_{\max}(d) = \frac{1}{2}+\frac{d-1}{2d-1}.   \label{p_discrimination}
\end{eqnarray}

We note that the mean distance $<D(\rho_1,\rho_2)>$ and consequently 
the mean identification probability $p_{\max}(d)$ approach unity as the 
dimension $d$ goes to infinity.

%%%%%%%%%%%%%%%%%%%%%%%%%%%%%%%%%%%%%%%%%%%%%%%%%%%%%%%%%%%

\section{Quantum state identification: case of $N=1$}
Let us consider the simple case, $N=1$, where only one copy of each reference 
state is given.
We have three systems numbered by 0,1, and 2, each on $\BC^d$.
The input state $\rho$ is assumed to be given in system 0 and the reference 
states $\rho_1$ and $\rho_2$ in system 1 and system 2, respectively.
We specify the system that an operator acts on by a number in the parentheses;
$\rho(0)$ is a density operator on system 0 for example.
Our task is then to distinguish two states, 
     $\rho_1(0)\rho_1(1)\rho_2(2)$ and  $\rho_2(0)\rho_1(1)\rho_2(2)$ 
, which are assumed to occur with equal probabilities.
The POVM $\{E_1,E_2\}$ is a set of operators on the total system 
$0 \otimes 1 \otimes 2$ and should be 
independent of $\rho_1$ and $\rho_2$, since no classical knowledge 
on $\rho_1$ or $\rho_2$ is available in the state identification problem 
considered in this paper.

The averaged success probability of state identification is then expressed 
as
\begin{eqnarray}
  p^{(N=1)}(d)=\frac{1}{2}\sum_{a=1,2} \Big<
     \tr{E_a \rho_a(0) \rho_1(1) \rho_2(2)} \Big>,
\end{eqnarray}
where the average $<\cdots>$ should be performed by the unitary invariant and 
independent distribution of the two reference states with 
the integration measure defined in Eq.(\ref{measure}). 

Here it is very helpful to use the following formula for the unitary average 
of the tensor product of $n$ identically prepared pure states 
\cite{Hayashi04}:
\begin{eqnarray}
    < \rho^{\otimes n} > = \frac{\CS_n}{d_n}, \label{formula1}
\end{eqnarray}
where $\CS_n$ is the projector onto the totally symmetric subspace of 
$\{\BC^d\}^{\otimes n}$ and the dimension of the subspace is given by 
$d_n = \tr{\CS_n} = {}_{n+d-1}C_{d-1}$.
This relation can be derived either by an explicit computation with the 
integration measure (\ref{measure}) or more simply by the following group 
theoretical argument. 
For any unitary $U$ on $\BC^d$, we have 
$   \left< U^{\otimes n} \rho^{\otimes n} U^{+\otimes n} \right>
  =\left< \rho^{\otimes n} \right>, $
implying that the operator $\left< \rho^{\otimes n} \right>$ 
on the totally symmetric subspace 
of $\{\BC^d\}^{\otimes n}$ commutes with $U^{\otimes n}$ for any $U$. 
Shur's lemma then requires that $\left< \rho^{\otimes n} \right>$ be 
proportional to $\CS_n$, 
since $U^{\otimes n}$ acts on the totally symmetric space irreducibly.
The proportional coefficient turns out to be $1/d_n$ by a 
trace argument. Thus we obtain formula (\ref{formula1}).

Performing the average integration by the use of formula (\ref{formula1}) 
and using the completeness of POVM, $E_1+E_2=1$, we find
\begin{eqnarray}
  p^{(N=1)}(d) = \frac{1}{2}+\frac{1}{2d_2d_1}\tr{E_1(\CS_2(01)-\CS_2(02))},
\end{eqnarray}
where $\CS_2(01)$ and $\CS_2(02)$ are the projectors onto the totally 
symmetric subspace of space $0 \otimes 1$ and $0 \otimes 2$, respectively. 

For any Hermitian traceless operator $D$, we observe the relation
\begin{eqnarray}
  \max_{0 \le E \le 1} \tr{ED} = \frac{1}{2}{\rm tr}|D|.  \label{D_inequality}
\end{eqnarray}
Here the maximum is attained if and only if $E=\CP_{D_+} + E_{D_0}$, 
where $\CP_{D_+}$ is the projector onto the eigenspace with positive 
eigenvalues of $D$, and $E_{D_0}$ is any operator in the subspace 
of zero-eigenvalue of $D$ with the condition $0 \le E_{D_0} \le 1$. 
Note that the operator $E$ can be chosen as a projector by taking 
$E_{D_0}=\CP_0$, where $\CP_0$ is is a projector onto a subspace spanned by 
some zero-eigenvalue states of $D$, 
which leads to a projective optimal measurement.

Taking $D = \CS_2(01)-\CS_2(02)$, we obtain 
\begin{eqnarray}
        p^{(N=1)}_{\max}(d) = \frac{1}{2} + \frac{1}{4d_2d_1}
                                 {\rm tr}|\CS_2(01)-\CS_2(02)|. 
\end{eqnarray}
The optimal POVM can be given by the projective measurement, 
$E_1 = \CP_{D_+} + \CP_0$ and $E_2=1-E_1$.

In order to evaluate $p^{(N=1)}_{\max}(d)$ further, we must determine non-zero 
eigenvalues and their multiplicities of the operator 
$D \equiv \CS_2(01)-\CS_2(02) = \frac{1}{2} (T(01)-T(02))$, 
where $T(01)$ is the transposition operator between systems 0 and 1 
and $T(02)$ is defined similarly. 

It is instructive to explicitly work out the case of $d=2$ first.
It is easy to see that the totally symmetric states are eigenstates of 
$D$ with zero eigenvalue
\begin{eqnarray}
  && D \ket{000} = 0, \nonumber \\
  && D \ket{111} = 0, \nonumber \\
  && D (\ket{100}+\ket{010}+\ket{001}) = 0, \nonumber \\
  && D (\ket{011}+\ket{101}+\ket{110}) = 0.
\end{eqnarray}
For mixed symmetric states, we find four eigenvectors in the following way.
\begin{eqnarray}
   D \ket{\pm:100} &=& \pm\frac{\sqrt{3}}{2} \ket{\pm:100},
                               \nonumber \\
   D \ket{\pm:110} &=& \pm\frac{\sqrt{3}}{2} \ket{\pm:110},
\end{eqnarray}
where
\begin{eqnarray}
 \ket{\pm:100} &=& \left(
       -2\ket{100}+(1\mp\sqrt{3})\ket{010}+(1\pm\sqrt{3})\ket{001}
                   \right)/\sqrt{12},
                          \nonumber \\
 \ket{\pm:110} &=& \left(
      -2\ket{011}+(1\mp\sqrt{3})\ket{101}+(1\pm\sqrt{3})\ket{110}
                   \right)/\sqrt{12}. 
\end{eqnarray} 
There is no totally antisymmetric state in this $d=2$ case. 
Thus eigenvalues of $D$ are four 0's, two $\frac{\sqrt{3}}{2}$'s and two 
$-\frac{\sqrt{3}}{2}$'s, which yields ${\rm tr}|D|=2\sqrt{3}$. 
The optimal measurement in this case is given by the following projective 
measurement: 
\begin{eqnarray}
  E_1 &=& \ket{+:100}\bra{+:100} + \ket{+:110}\bra{+:110} + \CP_0, 
                              \nonumber \\
  E_2 &=& \ket{-:100}\bra{-:100} + \ket{-:110}\bra{-:110} + \CS_3(012)-\CP_0, 
\end{eqnarray}
where $\CS_3(012)$ is the projector onto the totally symmetric space and 
$\CP_0$ is a projector onto any subspace of it.
One might wonder why the freedom in choosing $\CP_0$ does not affect 
the optimal success probability. The reason is that if the whole system 
is measured and found to be totally symmetric, we conclude that the system 
was prepared equally likely in one of the two states, 
$\rho_1(0)\rho_1(1)\rho_2(2)$ or $\rho_2(0)\rho_1(1)\rho_2(2)$, 
and any guess does not improve the success probability. 

Rather surprisingly, this property of the spectrum of operator $D$ in 
the qubit case prevails in general dimensions $d$.
This can be most conveniently seen by computing $D^2$ as follows:
\begin{eqnarray}
    D^2 = \frac{1}{4}\Big( 2-T(01)T(02)-T(02)T(01) \Big)
        = \frac{3}{4}\Big( 1 - \CS_3(012)-\CA_3(012) \Big)
        = \frac{3}{4}\CM_3(012),
\end{eqnarray}
where $\CA_3(012)$ and $\CM_3(012)$ are the projectors onto the totally 
antisymmetric and the mixed symmetric subspace of space 
$0 \otimes 1 \otimes 2$, respectively. 
From this we immediately obtain $|D| = \frac{\sqrt{3}}{2}\CM_3(012)$, 
implying 
${\rm tr}|D| = \frac{1}{\sqrt{3}}d(d^2-1)$, 
since the dimension of the mixed symmetric subspace of $(\BC^d)^{\otimes 3}$ 
is given by $\frac{2}{3}d(d^2-1)$.

The maximal averaged identification probability $p_{\max}^{(N=1)}(d)$ for 
general $d$ is then given by
\begin{eqnarray}
  p_{\max}^{(N=1)}(d) = \frac{1}{2} + \frac{\sqrt{3}(d-1)}{6d},
\end{eqnarray}
which is certainly less than the averaged maximum discrimination probability 
$p_{\max}(d)$ given in Eq.({\ref{p_discrimination}).

%%%%%%%%%%%%%%%%%%%%%%%%%%%%%%%%%%%%%%%%%%%%%%%%%%%%%%%%%%%%%%

\section{Quantum state identification: case of arbitrary $N$}
The formulation proceeds in a similar way for arbitrary number ($N$) 
of copies of the reference states.
We assume $N$ copies of each reference state $\rho_a\ (a=1,2)$ are prepared 
in systems $a_1$ to $a_N$.  
We denote the subsystem $a_1 \otimes a_2 \otimes \cdots \otimes a_N$ simply by 
$a$ ($a=1,2$).
The POVM now acts on the total system 
$0 \otimes 1 \otimes 2 = (\BC^d)^{\otimes (2N+1)}$. 
Then the averaged success probability can be calculated in the same way 
as in the $N=1$ case as follows:
\begin{eqnarray}
  p^{(N)}(d) &=& \frac{1}{2} \sum_{a=1,2} 
      \Big< \tr{E_a\rho_a(0)\rho_1(1)^{\otimes N}\rho_2(2)^{\otimes N}} \Big>
                       \nonumber \\
   &=& \frac{1}{2} + \frac{1}{2d_{N+1}d_N} 
       \tr{E_1(\CS_{N+1}(01)\CS_N(2) - \CS_N(1)\CS_{N+1}(02))} 
                        \nonumber \\
  &\le&
          \frac{1}{2} + \frac{1}{4d_{N+1}d_N} 
       {\rm tr}\Big|\CS_{N+1}(01)\CS_N(2) - \CS_N(1)\CS_{N+1}(02)\Big|
                        \nonumber \\
  &\equiv&
          p^{(N)}_{\max}(d).
\end{eqnarray}

We note that it suffices to work in the subspace where systems 1 and 2 are 
both totally symmetric
in order to determine non-zero eigenvalues and their multiplicities of 
the operator, 
\begin{eqnarray}
    D \equiv \CS_{N+1}(01)\CS_N(2)-\CS_N(1)\CS_{N+1}(02).
\end{eqnarray} 
By the argument in the preceding section, we find the optimal POVM 
can be given by a projective measurement in this general case as well, 
$E_1 = \CP_{D_+} + \CP_0$ and $E_2=1-E_1$.

%%%%%%%%%%%%%%%%%%%%%%%%%%

\subsection{Case of qubits ($d=2$)}
In this case we find that the algebra of the angular momentum operators 
is useful. As stated above, we can work in the subspace where systems 1 and 
2 are both totally symmetric, namely the total angular momentum of each system 
is $\frac{N}{2}$. The problem is then reduced to the recoupling of 
three angular momenta $\frac{1}{2}$, $\frac{N}{2}$ and $\frac{N}{2}$. 

It should be noted that the support of the projector $S_{N+1}(0a)$ $(a=1,2)$ 
is the subspace where the two angular momenta of systems 0 and $a$ are coupled 
to the total angular momentum $\frac{N}{2}+\frac{1}{2}$.
Therefore, the projector $\CS_{N+1}(0a)$ can be expressed in 
terms the angular momentum operators as
\begin{eqnarray}
  \CS_{N+1}(0a) = \frac{1}{N+1} 
        \left( 2\bold{j}(a) \cdot \bold{s}(0) + \frac{N}{2} + 1 
                      \right),\ (a=1,2),  
\end{eqnarray}
where $\bold{s}(0)$ is the spin operator of system 0 and $\bold{j}(a)$ is 
the angular momentum operator of system $a(=1,2)$. 
We note that $\bold{j}(a)^2 = \frac{N}{2}(\frac{N}{2}+1)$.

Since $\CS_N(a)$ is the identity in the subspace under consideration,
we find 
\begin{eqnarray}
    D = \frac{2}{N+1} \Big( \bold{j}(1)-\bold{j}(2) \Big) \cdot\bold{s}(0).
\end{eqnarray}
Then $D^2$ can be readily calculated by use of the properties of the Pauli 
matrices and the commutation relations of angular momenta $\bold{j}(1)$ and 
$\bold{j}(2)$. The result is given by
\begin{eqnarray}
   D^2 = \frac{1}{(N+1)^2} \left(
             (N+\frac{1}{2})(N+\frac{3}{2}) - \bold{J}^2 \right),
                    \label{D2_qubit}
\end{eqnarray}
where $\bold{J} = \bold{s}(0) + \bold{j}(1) + \bold{j}(2)$ is the total 
angular momentum operator.
From this we obtain the non-zero eigenvalues of $D$ to be 
$\pm \delta^{(J)}$ with multiplicity $2J+1$, where $\delta^{(J)}$ is given by
\begin{eqnarray}
 \delta^{(J)} &=& \frac{1}{N+1} 
              \sqrt{ (N+\frac{1}{2})(N+\frac{3}{2}) - J(J+1)}
                         \nonumber \\
          &=& \sqrt{1-\left(\frac{J+\frac{1}{2}}{N+1}\right)^2},\ \ 
            ( J=\frac{1}{2},\frac{3}{2},\cdots,N-\frac{1}{2}).
                    \label{Delta_qubit}
\end{eqnarray}

Thus we find the mean identification probability 
for qubit case to be 
\begin{eqnarray}
  p^{(N)}_{\max}(d=2) = \frac{1}{2} + \frac{1}{2(N+1)(N+2)}
       \sum_{J=\frac{1}{2}}^{N-\frac{1}{2}} (2J+1)
          \sqrt{1-\left(\frac{J+\frac{1}{2}}{N+1}\right)^2 }.
                      \label{pmax2}
\end{eqnarray}

We list explicit values of $p^{(N)}_{\max}(d=2)$ for some small $N$'s.
\begin{eqnarray}
   p^{(1)}_{\max}(d=2) &=& \frac{1}{2}+\frac{1}{12}\sqrt{3}
                               \, \simeq\, 0.644,  \\
   p^{(2)}_{\max}(d=2) &=& \frac{1}{2}+\frac{1}{18}(\sqrt{2}+\sqrt{5}) 
                                        \,\simeq\, 0.703,   \\
   p^{(3)}_{\max}(d=2) &=& \frac{1}{2}+
                             \frac{1}{80}(\sqrt{15}+4\sqrt{3}+3\sqrt{7}) 
                           \,\simeq\, 0.734\ . 
\end{eqnarray}

The value of $p^{(N)}_{\max}(d=2)$ in large $N$ limit can be obtained 
by replacing the sum in (\ref{pmax2}) with a continuous integration.
\begin{eqnarray}
   p^{(N)}_{\max}(d=2) \rightarrow \frac{1}{2} + 
               \int_0^1\! dx \, x\sqrt{1-x^2} = \frac{5}{6},
               \ (N \rightarrow \infty),
\end{eqnarray}
which is equal to the mean discrimination probability $p_{\max}(d=2)$ 
as expected.

%%%%%%%%%%%%%%%%%%%%%%%%%%

\subsection{Case of arbitrary dimension $d$}
Now we study the case of general dimension, which is naturally more involved 
than the qubit case. 
Let us introduce the orthonormal base of the total system 
$(\BC^d)^{\otimes (2N+1)}$ according to irreducible representations 
of the symmetric group $S_{2N+1}$ and the unitary group $U(d)$:
\begin{eqnarray}
     \ket{\lambda,a,b}.
\end{eqnarray}
In this base, $\lambda$ represents an irreducible representation 
of $S_{2N+1}$, which is specified by a Young diagram. 
The expression $\lambda=[\lambda_1,\lambda_2,\cdots]$ means a 
Young diagram consisting of a set of rows with their lengths given 
by $\lambda_1,\lambda_2,\cdots$. 
The label $a$ indexes vectors in a particular $S_{2N+1}$ representation space 
and it runs from 1 to the dimension of the $S_{2N+1}$ representation.
The $\lambda$ also specifies irreducible representations of the unitary 
group $U(d)$ and its vectors are indexed by $b$, which runs from 1 to 
$m_\lambda(d)$, the multiplicity of representation $\lambda$ of 
$S_{2N+1}$ on $(\BC^d)^{\otimes (2N+1)}$ \cite{Hamermesh62}. 
We study the spectrum of $D$ in the base of $\ket{\lambda,a,b}$. 

First we observe that $D$ commutes with $U^{\otimes (2N+1)}$ for arbitrary 
unitary $U$, since $D$ involves only permutation operators among the $2N+1$ 
subsystems. By Shur's lemma, $D$ should be proportional to the identity in a 
particular irreducible representation space of $U(d)$. We can always 
choose the label $a$ so that $D$ is diagonal with respect to it.
Then we have
\begin{eqnarray}
     D \ket{\lambda,a,b} = \Delta^\lambda_a \ket{\lambda,a,b}.
\end{eqnarray}
where $\Delta^\lambda_a$ is an eigenvalue of $D$, which is independent of $b$.

We are interested only in non-zero eigenvalues of $D$, therefore, we work 
in the subspace, denoted by $V_s$, in which system 1 and 2 are both totally 
symmetric. The space $V_s$ is the product of three $U(d)$ 
irreducible representations $[1]\otimes[N]\otimes[N]$. 
Decomposing $V_s$ into irreducible representations of $U(d)$, we classify 
the base states in $V_s$ into three groups. 

The first is the totally 
symmetric state:
\begin{eqnarray}
    \ket{[2N+1],b},\ \ b=1,\cdots,m_{[2N+1]}(d),
\end{eqnarray}
whose eigenvalues of $D$ are obviously zero.
The states in the second group belong to representations specified by 
Young diagrams of two rows $[\lambda_1,\lambda_2]$, where 
$N+1 \le \lambda_1 \le 2N$ and $\lambda_1+\lambda_2=2N+1$. 
Since each of these $U(d)$ representations appears twice in $V_s$, 
we distinguish the two by label $a=\pm$ as follows:  
\begin{eqnarray}
   \ket{[\lambda_1,\lambda_2],a,b},\ \ 
    a=\pm,\ \  
    b=1,\cdots,m_{[\lambda_1,\lambda_2]}(d).
\end{eqnarray}
The remaining states are those whose Young diagram has three rows. 
We do not need the label $a$, since each of these representations 
occurs only once in $V_s$.
\begin{eqnarray}
   \ket{[\lambda_1,\lambda_2,\lambda_3],b},\ \ 
   b=1,\cdots,m_{[\lambda_1,\lambda_2,\lambda_3]}(d).
\end{eqnarray}

Now we introduce the exchange operator $T(12)$ between systems 1 and 2 by
\begin{eqnarray}
   T(12) \equiv \prod_{i=1}^N T(1_i2_i).
\end{eqnarray}
It is readily seen that $T(12)$ anticommutes with $D$, implying that 
if $\ket{\phi}$ is an eigenstate of $D$, then $T(12)\ket{\phi}$ is also 
an eigenstate with the eigenvalue of opposite sign. On the other hand, 
the operation of $T(12)$ does not change quantum numbers $\lambda$ or 
$b$, since $T(12)$ is just a permutation operator. 
Thus we conclude that the states with three rows have eigenvalue zero. 
And the label $a=\pm$ for the states with two rows can be chosen so that 
$\Delta^\lambda_+ \ge 0$ and $\Delta^\lambda_-=-\Delta^\lambda_+$.
Therefore, ${\rm tr}|D|$ is given by
\begin{eqnarray}
    {\rm tr}|D| = 2\sum_{\lambda_1=N+1}^{2N} m_{[\lambda_1,\lambda_2]}(d)
                        \Delta^{[\lambda_1,\lambda_2]}_+.
\end{eqnarray}

We must still evaluate $\Delta^{[\lambda_1,\lambda_2]}_+$. 
However, it should be noticed that this is independent of $d$ and 
without loss of generality we can assume $d=2$ in the calculation of 
$\Delta^{[\lambda_1,\lambda_2]}_+$. 
This has already been calculated and given in Eq.(\ref{Delta_qubit}) in 
the preceding subsection. In the case of $d=2$, the total angular momentum 
$J$ is given by $J=\frac{\lambda_1-\lambda_2}{2}= \lambda_1-N-\frac{1}{2}$.
We find 
\begin{eqnarray}
   \Delta^{[\lambda_1,\lambda_2]}_+ = \delta^{(J = \lambda_1-N-\frac{1}{2})}
   =  \sqrt{1- \left( \frac{\lambda_1-N}{N+1} \right)^2}.
\end{eqnarray}

Thus we obtain the averaged success probability to be 
\begin{eqnarray}
  p^{(N)}_{\max}(d) = \frac{1}{2} + \frac{1}{2d_{N+1}d_N}
     \sum_{\lambda_1=N+1}^{2N} m_{[\lambda_1,\lambda_2]}(d)
        \sqrt{1- \left( \frac{\lambda_1-N}{N+1} \right)^2},
\end{eqnarray}
where $m_{[\lambda_1,\lambda_2]}(d), (\lambda_2=2N+1-\lambda_1)$ is 
the multiplicity of the $S_{2N+1}$ irreducible 
representation $[\lambda_1,\lambda_2]$ in the total system 
$0 \otimes 1 \otimes 2$, 
\begin{eqnarray}
  m_{[\lambda_1,\lambda_2]}(d) = \frac{(\lambda_1+d-1)!(\lambda_2+d-2)!
                                       (\lambda_1-\lambda_2+1)}
                          {(d-1)!(d-2)!(\lambda_1+1)!\lambda_2!}.
\end{eqnarray}
The optimal projective POVM can be expressed in terms of the states 
$\ket{\lambda,a,b}$ as follows:
\begin{eqnarray}
  E_1 &=& \sum_{\lambda_1=N+1}^{2N} \ket{[\lambda_1,\lambda_2],+,b}
                                  \bra{[\lambda_1,\lambda_2],+,b}
              + \CP_0,       
                               \nonumber \\
  E_2 &=& 1 - E_1,
\end{eqnarray} 
where $P_0$ is a projector onto any subspace spanned by zero-eigenvalue 
states of $D$, $\ket{[2N+1],b}$ 
and $\ket{[\lambda_1,\lambda_2,\lambda_3],b}$.

In the large $N$ limit, we can verify that $p^{(N)}_{\max}(d)$ approaches 
the averaged maximum discrimination probability $p_{\max}(d)$ as follows:
\begin{eqnarray}
  p^{(N)}_{\max}(d) \rightarrow \frac{1}{2} + 
         (d-1)\!\!\int_0^1\!\!\!dx\, x (1-x^2)^{d-\frac{3}{2}} = \frac{1}{2} 
        + \frac{d-1}{2d-1} = p_{\max}(d) ,\ (N \rightarrow \infty).
\end{eqnarray}

%%%%%%%%%%%%%%%%%%%%%%%%%%%%%%%%%%%%%%%%%%%%%%%%%%%%%%%%%%%%%%%%%%%%%%%%

\section{Concluding remarks}
We have discussed a natural generalization of pure state discrimination 
problem, which we call pure state identification, 
where our task is to identify a given input state with one of the two 
reference states which are presented as its $N$ copies, but without classical 
information on them. 
The two reference states are assumed to be distributed in a unitary invariant 
way in the pure state space of dimension $d$. 
For arbitrary $N$ and $d$, we have determined the averaged maximal success 
probability, which was shown to be achievable by a projective measurement.
And it was explicitly shown that this mean identification probability 
approaches the mean discrimination probability in the large $N$ limit 
as expected.

In this paper, we assumed that the input state is given as a single copy.
If $M$ copies of the input state are given, our task is to distinguish 
two states $\rho_1^{\otimes M}(0)\rho_1^{\otimes N}(1)\rho_2^{\otimes N}(2)$ 
and $\rho_2^{\otimes M}(0)\rho_1^{\otimes N}(1)\rho_2^{\otimes N}(2)$.
By a similar argument to the one in the case of $M=1$, the averaged maximum 
success probability can be shown to take the form:
\begin{eqnarray}
   p_{\max}^{(M,N)}(d) = \frac{1}{2} + 
      \frac{1}{4d_{N+M}d_N} {\rm tr} 
         \Big| \CS_{N+M}(01)\CS_N(2) - \CS_N(1)\CS_{N+M}(02) \Big|.
\end{eqnarray}
For general $M$,$N$, and $d$, however, it does not seem straightforward to 
evaluate this probability further. 
We would like to leave this problem to a future study.

%%%%%%%%%%%%%%%%%%%%%%%%%%%%%%%%%%%%%%%%%%%%%%%%%%%%%%%%%%%%%%%%%%%%%%

\end{document}